\documentclass[aps,prb,preprint,showpacs,superscriptaddress,floatfix]{revtex4}
\usepackage{graphics}
\usepackage{epsfig}
\usepackage{amsmath}
\usepackage{amssymb}
\usepackage{calc}
\newcommand{\beq}{\begin{equation}}   
\newcommand{\eeq}{\end{equation}}   
\newcommand{\bea}{\begin{eqnarray}}   
\newcommand{\eea}{\end{eqnarray}}   

\begin{document}      
\title {Stringy Black Holes and the Geometry of Entanglement}
\author{P\'eter L\'evay} 
\affiliation{Department of Theoretical Physics, Institute of Physics, Technical University of Budapest, H-1521 Budapest, Hungary}  
\date{\today} 
\begin{abstract}
Recently striking multiple relations have been found between pure state $2$ and $3$-qubit entanglement  and extremal 
black holes in string theory.
Here we add further mathematical similarities which can be both useful in string and quantum information theory.
In particular we show that finding the frozen values of the moduli in the calculation of the macroscopic black hole entropy in the STU model, is related to finding the canonical form for a pure three qubit entangled state defined by the dyonic charges.
In this picture the extremization of the BPS mass with respect to moduli is connected to the problem of finding the optimal local distillation protocol of a GHZ state from an arbitrary three-qubit pure state.
These results and a geometric classification of STU black holes BPS and non-BPS can be described in the elegant language of twistors.
Finally an interesting connection between the black hole entropy and the average real entanglement of formation is established.
\end{abstract}      
\maketitle{}

\section{Introduction}
Recently there has been much progress in seemingly two unrelated fields of theoretical physics. One of them is quantum information theory which concerns 
the study of quantum entanglement the "characteristic trait of quantum mechanics"\cite{Sch} and its possible applications such as quantum
teleportation \cite{Bennett}, quantum cryptography \cite{Ben2} and
more importantly quantum computing \cite{Nielsen}.
The other is the physics of stringy black holes which has provided spectacular results such as the black hole attractor mechanism\cite{Ferrara} and the
microscopic calculation of the black hole entropy\cite{Strominger}
related to the nonperturbative symmetries found between different string theories \cite{Hull,Witten,Polchinski}.

As far as mathematics is concerned these two different strains of knowledge have turned out to be related when Duff\cite{Duff} pointed out that the  entropy of the so called extremal BPS STU black hole can be expressed in a very compact way in terms of Cayley's hyperdeterminant\cite{Gelfand} which plays a prominent role as the three-tangle\cite{ckw} in studies of three-qubit entanglement.
Recently further mathematical similarities have been found by Kallosh and Linde\cite{Linde}. They have shown that the entropy of the axion-dilaton black hole is related to the concurrence which is the unique pure two-qubit entanglement measure. They have streched the validity of the relationship between the three-tangle and the STU black hole entropy found by Duff to non-BPS black holes.
They have also related the well-known entanglement classes of pure three-qubit entanglement to different classes of black holes in string theory.
Finally they emphasized the univeral role of the Cartan-Cremmer-Julia $E_{7(7)}$ invariant playing as the expression for the entropy of black holes and black rings in $N=8$ supergravity/M-theory.
By making use of the $SU(8)$ symmetry they have pointed out that the three-tangle shows up in this invariant too.

These results are intriguing mathematical connections arising from the similar symmetry properties of qubit systems and
the web of dualities in the STU model.
As far as classical supergravity is concerned the symmetry of the extremal STU model is $SL(2, {\bf R})^{\otimes 3}$, or taking into account quantum corrections and the quantized nature of electric and magnetic charges $SL(2,{\bf Z})^{\otimes 3}$. In string theory the latter symmetry group is also dictated by internal consistency.
In qubit systems on the other hand the symmetry group in question is the group of stochastic local operations and classical communication (SLOCC) which is 
$SL(2,{ \bf C})^{\otimes 3}$.
Hence the groups connected to dualities occurring in stringy black holes
are related to integers or at most to the real number system.
However, the power of entanglement is related to the special role played by complex numbers in quantum theory. This manifests itself at the level of three-partite protocols in the use of the larger group $SL(2,{\bf C})^{\otimes 3}$ (or more generally in $GL(2, {\bf C})^{\otimes 3}$), giving rise to interesting complex geometry\cite{Levay1} similar to the one found in twistor theory\cite{Penrose,Ward}.

In the treatments of Refs.\cite{Duff,Linde} instead of the $8$ complex numbers characterizing a general (unnormalized) three-qubit state the $8$ integers corresponding to the quantized electric and magnetic charges of $N=2$ , $D=4$ 
supergravity has been used. Hence in this case we have a correspondence between
quantized charges and the integer amplitudes of a special class of (unnormalized) real three-qubit states.
Already using these real quantum bits or rebits\cite{Caves} enabled the authors of Refs.\cite{Duff,Linde} to obtain amazing formal correspondences between stringy black holes and quantum entanglement. 
Now the question arises: can we find further relationships displaying the power of three-qubit entanglement in the more general {\it complex} context?
One of the aims of the present paper is to answer this question in the affirmative.
We show that
the well-known process of finding the frozen values of the moduli for the calculation
of the macroscopic black hole entropy in the STU model,
is related to the problem of obtaining the canonical decomposition for the three-qubit states defined by the charges using {\it complex} SLOCC transformations.
We also regard this paper as an attempt to establish some sort of dictionary between the languages used by string theorists and researchers working in the field of quantum information theory.
In particular we would like to show how the general 
theory of {\it complex} three-qubit entanglement contains in the form of real states the important cases studied by string theorists in the special case of STU black holes.

The organization of the paper is as follows. In Section II. the background concerning three-qubit entanglement is presented. 
Here we also discuss the canonical form of three-qubit states (the analogue of the Schmidt decomposition for two-qubits), and its relationship to three-qubit invariants. 
Special attention is paid to the important special case of real states which will play the dominant role in subsequent chapters as the ones describing STU      black holes. Here a new result concerning a geometric characterization of       such real states embedded in the more general complex ones is obtained.
In section III. in the context of the supersymmetric STU-model we present the
quantum entanglement version of the well-known process of freezing the moduli by extremization of the BPS mass\cite{Ferrara} .
It turns out that this extremization is related to finding the optimal distillation protocol of a GHZ state in the entanglement picture.
The solution of the stabilization equations resulting in the STU black hole entropy formula have been obtained in \cite{Shmakova}. We show that the process of finding the frozen values of the moduli
is just the one of obtaining a canonical form
for the corresponding three-qubit state by employing {\it complex} SLOCC transformations. 
In Section IV.
using the complex principal null directions of the two-plane in ${\bf C}^4$ containing the two real four-vectors
of the charges  
we shed some light on the geometric meaning of this canonical form.
Here an alternative geometric picture for the classification of BPS and non-BPS black holes small and large is also suggested. It is based on the intersection properties of a complex line  in ${\bf CP}^3$ with a fixed quadric.   
Finally an interesting connection between the black hole entropy and the real entanglement of formation is established. This section also contains some comments and the conclusions.

\section{Three-qubit entanglement}

An arbitrary (unnormalized) three-qubit pure state $\vert \psi\rangle\in {\bf C}^2\otimes {\bf C}^2\otimes{\bf C}^2$ is characterized by
$8$ complex numbers ${\psi}_{lkj}$ with $l,k,j=0,1$ and can be written in the following form

\beq
\vert\psi\rangle =\sum_{l,k,j}{\psi}_{lkj}\vert lkj\rangle\quad
\vert lkj\rangle\equiv {\vert l\rangle}_C\otimes{\vert k\rangle}_B\otimes
{\vert j\rangle}_A.
\label{state}
\eeq
We can imagine three parties (Alice, Bob and Charlie), wildly separated,
possessing a qubit from the entangled three-qubit state $\vert\psi\rangle$.
(Above we have adopted the convention of Ref. [13] of labelling these qubits from the
right to the left.)
In a class of quantum information protocols the parties can manipulate their qubits {\it reversibly} with some probability of success
by performing {\it local} manipulations assisted by classical communication
between them. Such protocols are yielding special transformations of the states,
called stochastic local operations and classical communication (SLOCC).
It can be shown\cite{Vidal} that such operations can be represented mathematically by applying 
the group $GL(2,{\bf C})^{\otimes 3}$ on the state $\vert\psi\rangle$ in the form
\beq
\vert\psi\rangle\mapsto ({\cal C}\otimes{\cal B}\otimes{\cal A})\vert\psi\rangle, \quad {\cal C}\otimes{\cal B}\otimes{\cal A}\in GL(2, {\bf C})^{\otimes 3}.
\label{SLOCC}
\eeq
Since we are interested in states up to a physically irrelevant complex constant we can fix the determinants of the $GL(2, {\bf C})$ transformations to one, hence we can assume that the group of SLOCC transformations is just $SL(2, {\bf C})^{\otimes 3}$. 

In the SLOCC classification of pure three-qubit states one forms the space of equivalence classes ${\bf C}^2\otimes{\bf C}^2\otimes{\bf C}^2/ SL(2, {\bf C})^{\otimes 3}$.
The result is as follows\cite{Vidal}. We have six different equivalence classes. Four of them  correspond to the completely separable class $(A)(B)(C)$ represented e.g. by $\vert 000\rangle$, and three classes of biseparable states of the form
$A(BC)$, $B(AC)$ and $C(AB)$ represented e.g. by $(\vert 00\rangle +\vert 11\rangle)\otimes\vert 0\rangle$ for the first of them.
The remaining two classes are the so-called Werner and Greenberger-Horne-Zeilinger classes represented by the states
$\vert W\rangle=\vert 001\rangle +\vert 010\rangle +\vert 100\rangle$
and $\vert GHZ\rangle=\vert 000\rangle +\vert 111\rangle$.
Hence apart from the separable cases three qubits can be entangled in two essentially different ways. The class carrying the genuine tripartite entanglement is the GHZ class. It is known that the GHZ state appears as the maximally entangled state\cite{Gisin}, it violates Bell-inequalities maximally and it maximizes the mutual information of local measurements, moreover it is the only state from which an EPR state can be obtained with certainty. 
On the other hand the W-state maximizes only two-qubit quantum correlations\cite{Vidal} inside our three-qubit state.

There are a number of polynomial invariants characterizing these entanglement classes. The most important one is the $SL(2, {\bf C})^{\otimes 3}$ and permutation (triality) invariant three-tangle\cite{ckw}
${\tau}_{ABC}\equiv 4\vert D({\psi})\vert$ where 
\begin{eqnarray}                                                                 D({\psi})&\equiv &  {\psi}_{000}^2{\psi}_{111}^2+{\psi}_{001}^2{\psi}_{110}^2+{\psi}_{010}^2{\psi}_{101}^2+{\psi}_{011}^2{\psi}_{100}^2\nonumber\\                  &-&2({\psi}_{000}{\psi}_{001}{\psi}_{110}{\psi}_{111}+{\psi}_{000}{\psi}_{010}{\psi}_{101}{\psi}_{111}\nonumber\\&+&                                            {\psi}_{000}{\psi}_{011}{\psi}_{100}{\psi}_{111}                                +{\psi}_{001}{\psi}_{010}{\psi}_{101}{\psi}_{110}\nonumber\\&+&                  {\psi}_{001}{\psi}_{011}{\psi}_{110}{\psi}_{100}+{\psi}_{010}{\psi}_{011}{\psi}_{101}{\psi}_{100})\nonumber\\ &4&({\psi}_{000}{\psi}_{011}{\psi}_{101}{\psi}_{110}+{\psi}_{001}{\psi}_{010}{\psi}_{100}{\psi}_{111})                            \label{hypdet}                                                                  \end{eqnarray}                                                                 \noindent                                                                       is the Cayley hyperdeterminant\cite{Gelfand}.
By chosing the first, second or third qubit one can introduce three sets of complex four vectors, e.g. by chosing the first i.e. Alice's qubit 
we define
\beq
{\xi}^{(A)}_I=\begin{pmatrix} {\psi}_{000}\\{\psi}_{010}\\{\psi}_{100}\\{\psi}_{110}\end{pmatrix},
\quad
{\eta}^{(A)}_J=\begin{pmatrix} {\psi}_{001}\\{\psi}_{011}\\{\psi}_{101}\\{\psi}_{111}\end{pmatrix}\quad
I,J=1,\dots 4
\label{A}
\eeq
similarly we can define the pairs of four-vectors $({\xi}^{(B)}, {\eta}^{(B)})$ and $({\xi}^{(C)}, {\eta}^{(C)})$.
Alternatively one can define three bivectors $P^{(A)}={\xi}^{(A)}\wedge {\eta}^{(A)}$
with components (Pl\"ucker coordinates)
\beq
P^{(A)}_{IJ}={\xi}^{(A)}_I{\eta}^{(A)}_J-{\xi}^{(A)}_J{\eta}^{(A)}_I
\label{Plucker}
\eeq
and similarly with the label $A$ replaced by $B$ or $C$.
Then we have\cite{Levay1, Levay2}
\beq
{\tau}_{ABC}=2\vert P^{(A)}_{IJ}P^{(A)IJ}\vert=
2\vert P^{(B)}_{IJ}P^{(B)IJ}\vert=
2\vert P^{(C)}_{IJ}P^{(C)IJ}\vert,
\label{invalak}
\eeq
where
indices are raised with respect to the $SL(2, {\bf C})\times SL(2, {\bf C})$ invariant metric $g={\varepsilon}\otimes {\varepsilon}$ 
\beq
\label{expl}
g^{IJ}=\begin {pmatrix} 0&0&0&1\\0&0&-1&0\\0&-1&0&0\\1&0&0&0\end {pmatrix}=
\begin {pmatrix} 0&1\\-1&0\end {pmatrix}\otimes
\begin {pmatrix} 0&1\\-1&0\end {pmatrix}.
\eeq
Since the Pl\"ucker coordinates (\ref{Plucker}) are $SL(2,{\bf C})$ invariant Eq. (\ref{invalak}) shows the $SL(2, {\bf C})^{\otimes 3}$ 
invariance and triality at the same time.
Notice that the three-tangle can also be written in the form
\beq
\label{alternative}
{\tau}_{ABC}=4\vert ({\xi}\cdot {\xi})({\eta}\cdot {\eta})-({\xi}\cdot {\eta})^2\vert =4\vert -D(\psi)\vert,
\eeq
with ${\xi}\cdot {\eta}\equiv g^{IJ}{\xi}_I{\eta}_J$ and
the possible labels $A,B,C$ of ${\xi}$ and ${\eta}$ are now supressed.

The physical importance of the three-tangle ${\tau}_{ABC}$ is that it discriminates between the two different types of three-qubit entanglement.
For the W-class we have ${\tau}_{ABC}=0$ and for the GHZ-class ${\tau}_{ABC}\neq 0$. In order to also discriminate between different types of separability we need further invariants.  
These are defined as follows.

Let us define the one and two partite reduced density matrices 
\beq
\label{reduced}
{\rho}_{A}={\rm Tr}_{BC}\vert\psi\rangle\langle\psi\vert,
\quad
{\rho}_{BC}={\rm Tr}_{A}\vert\psi\rangle\langle\psi\vert,
\eeq
and the quantities ${\rho}_{B}, {\rho}_{C}, {\rho}_{AC}$ and ${\rho}_{AB}$ are defined accordingly.
Note, that the trace of these quantities for unnormalized pure three-qubit states is not fixed to one.
Then we can define the quantity ${\tau}_{A(BC)}$ called the {\it squared-concurrence} between
the subsystems $A$ and $BC$ as
\beq
\label{conc1}
{\tau}_{A(BC)}=4{\rm Det}{\rho}_A=2\sum_{I,J=1}^4 {\overline P}^{(A)}_{IJ}P^{(A)}_{IJ}.
\eeq
Similarly one can define ${\tau}_{B(AC)}$ and ${\tau}_{C(AB)}$, using
$P^{(B)}$ and $P^{(C)}$ respectively.
Notice that now we have complex conjugation in the first factor and now the indices are {\it not} contracted by the metric $g$.
In order to understand this by supressing subsystem labels we can alternatively write
\beq
\label{conc2}
{\tau}_{A(BC)}=4(\langle {\xi}\vert {\xi}\rangle\langle {\eta}\vert {\eta}\rangle -{\vert\langle {\xi}\vert {\eta}\rangle\vert}^2),
\eeq
where $\langle  {\xi}\vert {\eta}\rangle =\sum_{\mu =1}^4\overline{\xi}_I{\eta}_I$. Eq. (11) 
should be compared with Eq.(\ref{alternative}).
We remark that the expressions for ${\tau}_{ABC}$ and ${\tau}_{A(BC)}$
can be written in a unified way by going to the "magic base"\cite{Wootters} via using a suitable unitary transformation.
In this case\cite{Levay1}  ${\tau}_{ABC}=2\vert P_{MN}P^{MN}\vert$
and ${\tau}_{A(BC)}=2\overline{P}_{MN}P^{MN}$ where
in this base indices are simply raised by ${\delta}^{MN}$  $M, N=1,2,3,4$ with $M$ and $N$ now labelling the components in the "magic base".
This way of writing uses the fact that $(SL(2, {\bf C})\otimes SL(2, {\bf C}))/{\bf Z}_2\simeq O(4,{\bf C})$. 
Here, in order to  establish connections with the formalism of stringy black holes, however, we follow a different route and use the somewhat more complicated expressions of Eqs. (\ref{invalak}), and (\ref{conc1}).
Notice also the factors of $2$ appearing in these formulae.
These are necessary for normalized states, since in this case all four quantities take values in the interval $[0,1]$.

Looking at Eq. (\ref{conc2}) it is clear that ${\tau}_{A(BC)}=0$ if and only if
${\xi}^{(A)}$ and ${\eta}^{(A)}$  are linearly dependent. (We exclude the trivial cases with $\xi$ or $\eta$ vanishing.)
This means that the corresponding reduced density matrix ${\rho}_A$
has rank one a condition equivalent to  $A(BC)$ separability.
Hence ${\tau}_{A(BC)}=0$ iff $\vert\psi\rangle$ is $A(BC)$ separable.
Similarly the vanishing of the squared concurrences ${\tau}_{B(AC)}$ and ${\tau}_{C(AB)}$ indicate separability of the form $B(AC)$ and $C(AB)$.

What about the invariance properties of our quantities ${\tau}_{A(BC)}$
,${\tau}_{B(AC)}$ and ${\tau}_{C(AB)}$? Clearly these quantities
are individually invariant with respect to $SL(2, {\bf C})\otimes U(4)$,
where the $SL(2, {\bf C})$ part is acting on the qubit which can be separated from the rest.
However, all three quantities are left invariant merely with respect to the action of the subgroup $SU(2)^{\otimes 3}$.

Using the four invariants ${\tau}_{ABC}$, ${\tau}_{A(BC)}$, ${\tau}_{B(AC)}$ and
${\tau}_{C(AB)}$ one can obtain the classification of pure three-qubit states\cite{Vidal}. For the completely separable class all of our invariants are vanishing.
 For the $A(BC)$ class only ${\tau}_{ABC}$ and ${\tau}_{A(BC)}$ is vanishing.
After the appropriate permutations the same can be said for the remaining biseparable classes. For the W-class only ${\tau}_{ABC}$ is vanishing, and at last for the GHZ-class none of the invariants is vanishing. 

How can we characterize two-partite correlations inside our three-qubit state?
In order to do this we have to look at the density matrices ${\rho}_{AB}$,
${\rho}_{AC}$ and ${\rho}_{BC}$.
Generally these states are mixed, so we have to characterize also  two-qubit mixed-state entanglement. A useful measure for the most general type of two-qubit mixed-state entanglement is\cite{Wootters}  ${\tau}_{AB}$ which is the {\it squared-concurrence for the mixed state} in question
\beq
\label{concurrence1}
{\tau}_{AB}=\left({\rm max}\{{\lambda}_1-{\lambda}_2-{\lambda}_3-{\lambda}_4,0\}\right)^2
\eeq
\noindent
where
${\lambda}_i, i=1,2,3,4$ is the nonincreasing sequence of the square-roots of the eigenvalues
of the nonnegative matrix 
 
\beq
\label{concurrence2}
{\rho}\tilde{\rho}\equiv{\rho}({\varepsilon}\otimes{\varepsilon})\overline{\rho}({\varepsilon}\otimes{\varepsilon}).
\eeq
The quantities ${\tau}_{AC}$ and ${\tau}_{BC}$ are defined accordingly.
Notice that the trace of the matrix ${\rho}\tilde{\rho}$ due to the Hermiticity of ${\rho}$ is an $SL(2, {\bf C})\times SL(2, {\bf C})$ invariant, since it is of the form
\beq
{\rm Tr}({\rho}\tilde{\rho})={\rho}_{IJ}{\rho}^{IJ}.
\eeq
\noindent
Consequently the traces of all powers of the matrix ${\rho}\tilde{\rho}$
are also invariant with respect to this group. The result is that the quantities
${\tau}_{AB}$, ${\tau}_{BC}$ and ${\tau}_{AC}$ are $SL(2, {\bf C})\otimes SL(2, {\bf C})$ invariant too.

In the special case when the mixed two-qubit state sits inside the pure three-qubit state
we have e.g. 
\beq
\label{BCsur}
({\rho}_{BC})_{IJ}={\xi}^{(A)}_I\overline{\xi}^{(A)}_J
+{\eta}^{(A)}_I\overline{\eta}^{(A)}_J,
\eeq
i.e. all of our two-qubit mixed state density matrices have rank at most two.
This means that in the formula (\ref{concurrence1}) we have at most two nonzero eigenvalues ${\lambda}_1$ and ${\lambda}_2$.
The invariants discussed above are not independent, they are subject to the important relations\cite{ckw}

\beq
\label{monogamy}
{\tau}_{A(BC)}={\tau}_{AB}+{\tau}_{AC}+{\tau}_{ABC}
\eeq
with the two other ones can be obtained by cyclic permutations.
These relations implying that e.g. ${\tau}_{A(BC)}\leq{\tau}_{AB}+{\tau}_{AC}$ that are also called the entanglement monogamy relations expressing the fact that unlike classical, quantum correlations cannot be shared freely between the parties.

There is one more invariant whose geometric meaning was clarified in Ref.\cite{Levay1}.
Consider a pure three-qubit state which is nonseparable (i.e. none of the quantities ${\tau}_{A(BC)}$, ${\tau}_{B(AC)}$ and ${\tau}_{C(AB)}$ is vanishing.) Then the three separable bivectors $P={\xi}\wedge {\eta}$ (in the following the labels $A$, $B$ and $C$ are implicit, we refer to the triple of these objects by using plural for the corresponding quantities)  are giving rise to the planes
$a{\xi}_I+b{\eta}_I$ with $a,b\in{\bf C}$.
Then we can find the principal null directions
 of these planes by solving the quadratic equations $a^2({\xi}\cdot {\xi})+2ab
({\xi}\cdot {\eta})+b^2({\eta}\cdot {\eta})=0$. The discriminant
of these equations is just the Cayley hyperdeterminant so we have two principal null
directions for ${\tau}_{ABC}\neq 0$ and one  for ${\tau}_{ABC}=0$ for each plane.
 Hence the number of principal null directions corresponds to the two nonseparable three-qubit entanglement classes the W and the GHZ class.
 Assuming ${\xi}\cdot {\xi}\neq 0$ and solving the quadratic equations for the ratio $\frac
{a}{b}$,
 these directions are
 \beq
 \label{elso}
 u_I^{\pm}=-P_{IJ}{\xi}^J\pm \sqrt{D}{\xi}_I
,
 \eeq
 \noindent
 or alternatively assuming ${\eta}\cdot {\eta}\neq 0$ and solving for the ratio $\frac{b}{a}$
 
\beq
\label{masodik}
{v}_I^{\pm}=P_{IJ}{\eta}^J\pm \sqrt{D}{\eta}_I,
\eeq
\noindent
where
$D\equiv D(\psi)$ is the Cayley hyperdeterminant (\ref{hypdet}).
Of course these vectors are null i.e. $u^{\pm}\cdot u^{\pm}=v^{\pm}\cdot v^{\pm}=0$, moreover the two sets of solutions are proportional i.e. $u^{\pm}\sim v^{\mp}$.
One can show that
\beq
P_I^Ju^{\pm}_J=\mp \sqrt{D}u^{\pm}_J\quad
P_I^Jv^{\pm}_J=\pm \sqrt{D}v^{\pm}_J,
\eeq
\noindent
 i.e.they are eigenvectors of the Pl\"ucker matrix with eigenvalues $\pm 1$ times
the square root of Cayley's hyperdeterminant.

Let us now define the quantity
\beq
\label{sigma}
{\sigma}_{ABC}\equiv \vert\vert u^+\vert\vert^2+
\vert\vert u^-\vert\vert^2+
\vert\vert v^+\vert\vert^2+
\vert\vert v^-\vert\vert^2.
\eeq
It can be shown\cite{Levay1} that ${\sigma}_{ABC}$ is permutation and 
$SU(2)^{\otimes 3}$ invariant, and for normalized states takes values in the interval $[0,1]$.
(Remember that Eq. (\ref{sigma}) can be defined with three similar expressions with the corresponding quantities  $u^{\pm}$ and $v^{\pm}$ 
labelled by $A$ , $B$ and $C$. The three similar expressions turn out to be equal reflecting triality.) For the relationship of ${\sigma}_{ABC}$ to other permutation invariants expressed in terms of density operators see Refs. [14], [23].  

What is the significance of our new invariant ${\sigma}_{ABC}$?
It will turn out that the sufficient and necessary condition for an arbitrary {\it complex} three-qubit pure sate to be $SU(2)^{\otimes 3}$ equivalent to a {\it real} state can be expressed in terms of ${\sigma}_{ABC}$ in a simple form. These real states will be playing an important role in our description of stringy black holes in terms of three-qubit entanglement.

In order to find this condition we have to see how one can find canonical forms for three-qubit states\cite{Acin1,Acin2}.
For definiteness let us fix a qubit say $A$.
It was noted in\cite{Levay1} that finding this canonical form is equivalent to
first finding one of the principal null directions by performing a transformation $I\otimes I\otimes U_A$, with $U_A$ unitary and then performing further unitaries of the form $U_C\otimes U_B\otimes I$. 
After the first step we can have ${\xi}^{\prime}\cdot {\xi}^{\prime}=0$ , i.e.
${\rm Det}({\psi}^{\prime}_{lk0})=0$, and after the second ${\xi}^{\prime\prime}= (r_0,0,0,0)^T$,
with $r_0$ a real number. The result of this process for the canonical form is
\cite{Acin1,Acin2}
\beq
\label{canonical}
\vert\psi\rangle =r_0\vert 000\rangle +e^{i\varphi}r_1\vert 001\rangle +r_2\vert011\rangle +r_3\vert 101\rangle +r_4\vert 111\rangle,
\eeq
\noindent
where the numbers $r_a,\quad a=0,\dots 4$ are real nonnegative and $0\leq\varphi\leq\pi$.
Notice that unlike in the two qubit case where the canonical form (the well-known Schmidt decomposition) contains merely two real nonnegative numbers, here we
also have an unremovable complex phase. 
Note also that this decomposition is unique for $0< \varphi <\pi$.
For the remaining cases $\varphi=0,\pi$ two canonical forms exist (corresponding to the two principal null directions).
One can break this degeneracy by taking the form with the smallest value for $r_1$, or if $r_1$ is unique taking the form with the smallest $r_0$\cite{Acin2}.

Based on the results of Ref. [25] we can show that the expansion coefficients $r_a$, $a=0,1,\dots 4$ and $\cos\varphi$ can be expressed in terms of the invariants 
${\tau}_{AB}$, ${\tau}_{BC}$, ${\tau}_{AC}$, ${\tau}_{ABC}$ and ${\sigma}_{ABC}$. It is straightforward to show that Eqs. (24)-(27) of that paper in our notation look like
\beq
\label{r0}
(r_0^{\pm})^2=\frac{{\sigma}_{ABC}\pm\sqrt{\Delta}}{2({\tau}_{AB}+{\tau}_{ABC})}
\eeq
\beq
\label{r2}
(r_2^{\pm})^2=\frac{{\tau}_{AC}({\tau}_{AB}+{\tau}_{ABC})}{2({\sigma}_{ABC}\pm\sqrt{\Delta})}
\eeq
\beq
\label{r3}
(r_3^{\pm})^2=\frac{{\tau}_{BC}({\tau}_{AB}+{\tau}_{ABC})}{2({\sigma}_{ABC}\pm\sqrt{\Delta})}
\eeq
\beq
\label{r4}
(r_4^{\pm})^2=\frac{{\tau}_{ABC}({\tau}_{AB}+{\tau}_{ABC})}{2({\sigma}_{ABC}\pm\sqrt{\Delta})}
\eeq
\beq
\label{r1}
(r_1^{\pm})^2={\omega}_{ABC}-(r_0^{\pm})^2
-(r_2^{\pm})^2
-(r_3^{\pm})^2
-(r_4^{\pm})^2
\eeq
\beq
\label{fazis}
\cos{\varphi}^{\pm}=\frac{(r_1^{\pm}r_4^{\pm})^2+(r_2^{\pm}r_3^{\pm})^2-{\tau}_{BC}/4}{2r_1^{\pm}r_2^{\pm}r_3^{\pm}r_4^{\pm}},
\eeq
\noindent
where
\beq
\label{Delta}
{\Delta}={\sigma}_{ABC}^2-({\tau}_{AB}+{\tau}_{ABC})
({\tau}_{BC}+{\tau}_{ABC})({\tau}_{AC}+{\tau}_{ABC}).
\eeq
\noindent
Notice that due to the fact that our states are unnormalized the norm squared ${\omega}_{ABC}=\langle\psi\vert\psi\rangle$ as an obvious $SU(2)^{\otimes 3}$ invariant appears.

Now at last, how can we characterize real states inside the complex ones?
A pure three-qubit state is said to be real when there exists a product basis where all coefficients are real.
There is a theorem\cite{Acin2} stating that a pure three-qubit state is real
if and only if 
\beq
\label{cond1}
\sqrt{{\tau}_{AB}{\tau}_{BC}{\tau}_{AC}}=\vert {\sigma}_{ABC}-{\omega}_{ABC}{\tau}_{ABC}\vert
\eeq
\noindent
or
\beq
\label{cond2}
{\Delta}=0
\eeq
\noindent
holds.
Notice that unlike in Ref. [25] in these reality conditions as the first result of this paper the role of geometry via the occurrence of the principal null directions is clearly displayed. Actually in the paper of Acin et.al.\cite{Acin2} the reality conditions are not even expressed in terms of our fundamental invariants.  
Looking back to the quadratic equations determining the principal null directions one can show that if the initial states are real then (\ref{cond1}) holds and the null directions are both real, or (\ref{cond2}) holds  and the null directions are complex conjugate of each other.
In the second case from Eqs. (22)-(27) one can see that in this case  $r_a^+=r_a^-$ and $\cos{\varphi}^+=\cos{\varphi}^-$.
Notice the simple form of the coefficients $r_a$ for the ${\Delta}=0$ case. 
As we will see in the next section the ${\Delta}=0$ case will hold for supersymmetric BPS black holes, and the case characterized by Eq.(\ref{cond1}) will correspond to non-supersymmetric non-BPS black holes.

Closing this section we note the following important facts to be used later.
In order to reach a canonical form we can start by choosing any of the qubits to play a special role. In order to preserve the norm untill this point we used unitary transformations to obtain this canonical form.
However, for unnormalized states we can relax this constraint and we can use
the more general class of SLOCC transformations on the chosen qubit, while for the remaining ones we can continue using local unitaries. 
As we have seen this process will still result in a five term canonical form.
However, if we chose the full group of SLOCC transformations than we can reach
the simpler looking representative states of the SLOCC classes, namely the separable, biseparable, W and GHZ classes.
Starting from an arbitrary complex state for the special case with $D\neq 0$ we can arrive at the canonical $GHZ$ state $\vert 000\rangle +\vert 111\rangle$.
However, from the real states with $\Delta =0$ we can only reach $GHZ$ states
of the form\cite{Acin2}
\beq
\vert {\psi}^{\prime}\rangle=\alpha(\vert 000\rangle +e^{i\delta}\vert 111\rangle)
\label{slocckan}
\eeq
\noindent
with $\delta$ generally not equal to $0$.
This canonical form will play an important role in our later considerations concerning BPS STU black holes.
This completes our study of three-qubit entanglement of the most general {\it complex} type.
In the following section we turn our attention to a very special class of three qubit entanglement. Representative states will be unnormalized and having integer amplitudes.
These states and their complexifications will describe the entanglement properties of STU black holes.

\section{STU black holes and entanglement}

Based on the results of the previous section now we establish some new 
connections
between the theory of three-qubit quantum entanglement and the $STU$ model admitting extremal black hole solutions.
In the following we consider ungauged $N=2$ supergravity in $D=4$ coupled to
$n$ vector multiplets. At first the number $n$ will be arbitrary we will specialize to the $n=3$ case corresponding to the $STU$ model later.
The Lagrangian of such models can be constructed\cite{Witt} and the relevant piece of its bosonic part that we need is of the form\cite{Kallosh}
\begin{eqnarray}                                                               
\label{Lagrangian}
{\cal L}&=&\frac{1}{2}\int d^4x\sqrt{-g}\{-R+2G_{i\overline{j}}{\partial}_{\mu}z^i{\partial}_{\nu}{\overline{z}}^{\overline{j}}g^{\mu\nu}\nonumber\\&+&2({\rm Im}{\cal N}_{IJ}{\cal F}^I{\cal F}^J+{\rm Re}{\cal N}_{IJ}{\cal F}^I{^\ast{\cal F}^J})\}
\end{eqnarray}                                                                 
Here ${\cal F}^I$, and ${^\ast{\cal F}^I}$,  $I=1,2\dots n+1$ are two-forms associated to the field strengths ${\cal F}^I_{\mu\nu}$  of $n+1$ $U(1)$ gauge-fields and their duals. 
The $z^i$ $i=1,\dots n$ are complex scalar fields that can be regarded as local coordinates on a projective special K\"ahler manifold ${\cal M}$. This manifold can be defined by 
constructing a flat symplectic bundle of dimension $2n+2$ over a K\"ahler-Hodge manifold with a symplectic section
$(L^I(z,\overline{z}), M_J(z,\overline{z})),\quad I,J=1,\dots n+1$ satisfying
\beq
i(\overline{L}^I M_I-L^I\overline{M}_I)=1.
\eeq
\noindent 
Here
$L^I$ and $M_J$ are covariantly holomorphic with respect to the K\"ahler connection implying that after introducing the holomorphic sections $(X^I, F_J)$ as
\beq
L^I=e^{K/2}X^I,\quad M_J=e^{K/2}F_J,\quad {\partial}_{\overline{i}}X^I={\partial}_{\overline{j}}F_J=0
\eeq
the K\"ahler metric is $G_{i\overline{j}}={\partial}_i{\partial}_{\overline{j}}K$ with the K\"ahler potential 
\beq
\label{potencial}
K=-{\rm ln}i(\overline{X}^IF_I-X^I\overline{F}_I).
\eeq
\noindent
Finally the complex symmetric matrix ${\cal N}_{IJ}$
satisfies the constraints
\beq
\label{Nmatrix}
M_I={\cal N}_{IJ}L^J, \quad
{\rm Im}{\cal N}_{IJ}L^I\overline{L}^J=-\frac{1}{2}.
\eeq
\noindent
and
\beq
D_{\overline{i}}\overline{M}_I={\cal N}_{IJ}D_{\overline{i}}\overline{L}^J,
\quad D_{\overline{i}}={\partial}_{\overline{i}}-\frac{1}{2}K_{\overline{i}}.
\eeq
\noindent
For the physical motivation of Eq. (\ref{Lagrangian}) we note that such Lagrangians arise by dimensional reduction of the ten-dimensional string theory on a compact six dimensional manifold ${\cal K}$ and restriction to massless modes.
In this case our ${\cal M}$ is just the moduli space of ${\cal K}$.
Indeed, Calabi-Yau three-folds provide moduli spaces as realizations of special geometry\cite{Strominger2}. 

Defining
\beq
{\cal G}_I={\rm Re}{\cal N}_{IJ}{\cal F}^J-{\rm Im}{\cal N}_{IJ}{^{\ast}{\cal F}}^J
\eeq
\noindent
the covariant charges are defined as
\beq
\begin{pmatrix} p^I\\q_J\end{pmatrix}=\begin{pmatrix} \int{\cal F}^I\\\int{\cal
G}_J\end{pmatrix}.
\eeq
\noindent
The central charge formula is given by
\beq
Z(z,\overline{z},p,q)=e^{K(z,\overline{z})/2}(X^I(z)q_I-F_I(z)p^I).
\label{central}
\eeq
As we see the central charge is depending on the charges and the moduli $z^i$.
Note that $z^i$ are space-time dependent.
It is well-known that extremal BPS black hole solutions to the equations of motion corresponding to the Lagrangian (\ref{Lagrangian}) can be found. These are static, spherically symmetric, asymptotically flat solutions with regular event horizons.
The solutions contain besides the metric our $n+1$ gauge-fields and $n$ scalars $z^i$ both functions of the radial coordinate only.
Hence in these models the central charge (\ref{central}) is a function of the radial coordinate $r$.
In the asymptotically flat limit $r\to\infty$ we have
for the mass of the BPS black hole
\beq
M=\vert Z\vert_{\infty}=M(z^i(\infty), p,q),
\eeq
\noindent
i.e. it saturates the mass bound demanded by supersymmetry.
In the other (i.e. the near horizon) limit as in the case of the extremal Reissner-Nordstrom solution the metric takes the $AdS^2\times S^2$ form
\beq
\label{RN}
ds^2=-\frac{r^2}{{\vert Z\vert}^2_{\rm hor}}dt^2+\frac{{\vert Z\vert}^2_{\rm hor}}{r^2}dr^2+{\vert Z\vert}^2_{\rm hor}d{\Omega}^2
\eeq
\noindent
with ${\vert Z\vert}^2_{\rm hor}$ is the value of the central charge
at the horizon.
Since the area of the event horizon is $A=4\pi{\vert Z\vert}^2_{\rm hor}$ the macroscopic Bekenstein-Hawking entropy is
\beq
\label{entropy}
{\cal S}_{BH}=\frac{A}{4}=\pi{\vert Z\vert}^2_{\rm hor}.
\eeq
\noindent
${\cal S}_{BH}$ again seems to be depending on both the charges and the values of the moduli on the horizon. However, it turns out that the values
of the moduli on the horizon are determined by the charges\cite{Ferrara}.
This result is compatible with the one of relating ${\cal S}_{BH}$ a macroscopic entropy to a microscopic one which counts states\cite{Strominger}.
In string compactifications the fields $z^i(r)$ define a flow in moduli space
converging to a fixed point the "attractor"-value of the moduli determined by the charges.
The attractor equations equivalent to the ones coming from the extremization of the BPS mass
\beq
\label{BPS}
M^2_{BPS}={\vert Z\vert}^2=e^K{\vert X^Iq_I-F_Ip^I\vert}^2
\eeq
with respect to moduli
are of the form\cite{Ferrara}
\beq
\label{stabi}
\begin{pmatrix}p^I\\q_J\end{pmatrix}=2{\rm Im}\begin{pmatrix}Z\overline{L}^I\\Z\overline{M}_J\end{pmatrix}.
\eeq
\noindent
Equation (\ref{stabi}) provides a highly nontrivial constraint between the charges and the moduli.

There exist black-hole solutions for which the moduli remain constant even away from the horizon\cite{Shmakova,Kallosh}, hence in this case the black hole mass itself is also a function of the dyonic charges.
These solutions are called double extreme solutions.
In the following we will concentrate on such type of solutions. Moreover, in order
to find mathematical similarities with the three-qubit system we
restrict our attention to the $n=3$ case.
The double extreme solutions of the arising STU model were found by
Behrndt et. al.\cite{Shmakova} in the following we follow their notation.

For the STU-model we have $n=3$ 
and the corresponding three constant moduli are conventionally denoted as
$(z^1,z^2,z^3)=(S,T,U)$.
Our aim is to produce a quantum entanglement version of the determination of
their frozen value dictated by the supersymmetric attractor mechanism. 
We use special (inhomogeneous) coordinates for the holomorphic section $(X^I(z), F_J(z))$ as
\beq
\label{inhom}
X^I(z)=\begin{pmatrix}1\\z^1\\z^2\\z^3\end{pmatrix},\quad 
F_J(z)=\begin{pmatrix}-z^1z^2z^3\\z^2z^3\\z^1z^3\\z^1z^2\end{pmatrix}.
\eeq
\noindent
Recall that the model is described by the prepotential $F(X)=\frac{X^1X^2X^3}{X^0}$ i.e. $F_J=\frac{{\partial}F}{\partial X^J}$.
In accordance with Eq. (\ref{potencial}) the K\"ahler potential is
\beq
\label{STUpot}
K(z,\overline{z})=-\log(-i(z^1-\overline{z}^1)(z^2-\overline{z}^2)(z^3-\overline{z}^3)).
\eeq
\noindent
Then using the notation
\beq
S=S_1+iS_2,\quad T=T_1+iT_2,\quad U=U_1+iU_2
\eeq
\noindent
we can write  $M_{BPS}^2$ Eq. (\ref{BPS}) in the following form
\begin{widetext}
\beq
\label{BPSteljes}
M^2_{BPS}=\frac{1}{8S_2T_2U_2}{\vert q_0+q_1S+q_2T+q_3U+p^0STU-p^1TU-p^2SU-p^3ST\vert}^2.
\eeq
\end{widetext}
\noindent
We would like to write this expression in an alternative form reflecting triality\cite{Rahmfeld} 
\begin{widetext}
\beq
\label{Duff}
M^2_{BPS}=\frac{1}{8}{\psi}^T\left({\cal M}_U^{-1}\otimes{\cal M}_T^{-1}\otimes{\cal M}_S^{-1}-{\varepsilon}_U\otimes{\varepsilon}_T\otimes{\cal M}_S^{-1}-
		{\varepsilon}_U\otimes{\cal M}_T^{-1}\otimes{\varepsilon}_S-{\cal M}_U^{-1}\otimes{\varepsilon}_T\otimes{\varepsilon}_S\right){\psi},
		\eeq
		\end{widetext}
		\noindent
		where
		\beq
{\cal M}_S=\frac{1}{S_2}\begin{pmatrix}1&S_1\\S_1&\vert S\vert^2\end{pmatrix},
	\eeq
	\noindent
	with similar expressions for ${\cal M}_T$ and ${\cal M}_U$.
${\psi}$ is the hypermatrix of Eq. (\ref{state})
	defining our (real) three-qubit state.
	In order to find the exact relationship between the eight components of ${\psi}_{lkj}$ and the eight components of the two four vectors $(p^0,p^1,p^2,p^3)^t$
	and $(q_0,q_1,q_2,q_3)^t$ and to gain some additional insight we proceed as follows.

First let us write ${\cal M}_S^{-1}$ in the form
\beq
\label{split}
{\cal M}_S^{-1}={\cal A}^t_S{\cal A}_S,\quad {\cal A}_S=\frac{1}{\sqrt{S_2}}\begin{pmatrix}S_2&0\\-S_1&1\end{pmatrix},
\eeq
\noindent
similarly we define
\beq
\label{similar}
{\cal M}_T^{-1}={\cal B}^t_T{\cal B}_T\quad 
{\cal M}_U^{-1}={\cal C}^t_U{\cal C}_U,
\eeq
where the matrices ${\cal B}_T$ and ${\cal C}_U$ are defined accordingly.
Using ${\cal A}^t_S{\varepsilon}{\cal A}_S={\varepsilon}$ and similar expressions for ${\cal B}_T$ and ${\cal C}_U$ we get
\beq
\label{elsotrukk}
M^2_{BPS}={\psi}^t({\cal C}^t_U\otimes{\cal B}^t_T\otimes{\cal A}^t_S){\varrho}^{\prime}({\cal C}_U\otimes{\cal B}_T\otimes{\cal A}_S){\psi},
\eeq
\noindent
where
\beq
\label{surmatrix1}
{\varrho}^{\prime}\equiv\frac{1}{8}(I\otimes I\otimes I+{\sigma}_2\otimes{\sigma}_2\otimes I+{\sigma}_2\otimes I\otimes{\sigma}_2+I\otimes{\sigma}_2\otimes{\sigma}_2).
\eeq
\noindent
Notice that ${\varrho}^{\prime}=\frac{1}{2}{\Pi}$ where ${\Pi}$ is a rank-two projector, i.e. ${\Pi}^2={\Pi}$.
In other words ${\varrho}^{\prime}$ is a simple example of a mixed state three-qubit density matrix.
To reveal the rank-two structure of this density matrix we diagonalize ${\sigma}_2=-i\varepsilon$
\beq
\label{diag}
{\sigma}_3={\cal U}{\sigma}_2{\cal U}^{\dagger},\quad
{\cal U}=\frac{1}{\sqrt{2}}\begin{pmatrix}1&-i\\1&i\end{pmatrix}.
\eeq
\noindent
Then the new density matrix is
${\varrho}={\cal U}_U\otimes{\cal U}_T\otimes{\cal U}_S{\varrho}^{\prime}{\cal U}^{\dagger}_U\otimes{\cal U}^{\dagger}_T\otimes{\cal U}^{\dagger}_S=\frac{1}{2}{\rm diag}(1,0,0,0,0,0,0,1)$, or in the notation used in quantum information
\beq
\label{rho3}
{\varrho}=\frac{1}{2}(\vert 000\rangle\langle 000\vert +\vert 111\rangle\langle 111\vert ).
\eeq
\noindent
Using these unitary transformations we obtain a complex representation
for $M^2_{BPS}$ as follows
\beq
\label{nice}
M^2_{BPS}=\langle\psi\vert(C^{\dagger}_U\otimes B^{\dagger}_T\otimes A^{\dagger}_S){\varrho}(C_U\otimes B_T\otimes A_S)\vert\psi\rangle,
\eeq
\noindent
where $A_S$, $B_T$ and $C_U$ are now SLOCC i.e. $GL(2, {\bf C})$ transformations of the form
\beq
\label{SLOCC}
A_S=\frac{1}{\sqrt{2S_2}}\begin{pmatrix}\overline{S}&-1\\-S&1\end{pmatrix},
\eeq
\noindent
with $B_T$ and $C_U$ defined similarly.
Notice that we could have multiplied these $GL(2, {\bf C})$ matrices
by $e^{i\pi/4}$ rendering them to $SL(2,{\bf C})$ ones a transformation not changing $M^2_{BPS}$. 

Using the explicit form of $\varrho$ we obtain the nice result
\beq
\label{result}
M^2_{BPS}=\frac{1}{2}(\vert{\psi}^{\prime}_{000}\vert^2+\vert{\psi}^{\prime}_{111}\vert^2),
\eeq
\noindent
where 
\beq
\label{newstate}
\vert{\psi}^{\prime}\rangle\equiv (C_U\otimes B_T\otimes A_S)\vert \psi\rangle
\eeq
\noindent
is the SLOCC transformed state.
Looking at Eq.(\ref{result}) it is clear that $M^2_{BPS}$ is expressed in terms of the magnitudes of the GHZ part of the SLOCC transformed state depending on the values of the moduli $S$, $T$ and $U$ and their complex conjugates.

Choosing the first qubit as a reference (recall that we are labelling qubits from the right to the left) it is straightforward to show that
\begin{widetext}
\beq
\label{matr1}
\begin{pmatrix}{\psi}^{\prime}_{000}&{\psi}^{\prime}_{010}\\{\psi}^{\prime}_{100}&{\psi}^{\prime}_{110}\end{pmatrix}=\frac{1}{\sqrt{8S_2T_2U_2}}
\begin{pmatrix}\overline{U}&-1\\-U&1\end{pmatrix}
\begin{pmatrix}\overline{S}{\psi}_{000}-{\psi}_{001}&\overline{S}{\psi}_{010}-{\psi}_{011}\\\overline{S}{\psi}_{100}-{\psi}_{101}&\overline{S}{\psi}_{110}-{\psi}_{111}\end{pmatrix}
\begin{pmatrix}\overline{T}&-T\\-1&1\end{pmatrix},
\eeq
\end{widetext}
\begin{widetext}
\beq
\label{matr2}
\begin{pmatrix}{\psi}^{\prime}_{001}&{\psi}^{\prime}_{011}\\{\psi}^{\prime}_{101
}&{\psi}^{\prime}_{111}\end{pmatrix}=\frac{1}{\sqrt{8S_2T_2U_2}}
\begin{pmatrix}\overline{U}&-1\\-U&1\end{pmatrix}
\begin{pmatrix}-S{\psi}_{000}+{\psi}_{001}&-S{\psi}_{010}+{
\psi}_{011}\\-S{\psi}_{100}+{\psi}_{101}&-S{\psi}_{110}+{\psi
}_{111}\end{pmatrix}
\begin{pmatrix}\overline{T}&-T\\-1&1\end{pmatrix}.
\eeq
\end{widetext}
Calculating and substituting the components ${\psi}^{\prime}_{000}$ and ${\psi}^{\prime}_{111}$ into Eq. (\ref{result}) a comparison with
Eq.(\ref{BPSteljes}) yields the relation ${\psi}^{\prime}_{000}=-\overline{{\psi}^{\prime}}_{111}$ and the correspondence
\beq
\label{corres}
\begin{pmatrix}{\psi}_{000}\\{\psi}_{001}\\{\psi}_{010}\\{\psi}_{011}\end{pmatrix}=\begin{pmatrix}p^0\\p^1\\p^2\\q_3\end{pmatrix},\quad
\begin{pmatrix}{\psi}_{100}\\{\psi}_{101}\\{\psi}_{110}\\{\psi}_{111}           \end{pmatrix}
=\begin{pmatrix}p^3\\q_2\\q_1\\-q_0\end{pmatrix}.
\eeq
\noindent
Notice that our convention differs from the one of Duff\cite{Duff} in a sign change in the first four vector and from the one adopted by Kallosh and Linde\cite{Linde} by a change in sign of the components $q_0,p^1,p^2$ and $p^3$.

In order to proceed with the extremization of the BPS mass we introduce some notation. Let us label qubits instead of $A$, $B$ and $C$ by the letters $S$, $T$  and $U$. We still label qubits from the left to the right so for example
the four-vectors ${\xi}^{(S)}$ and ${\eta}^{(S)}$ are just the ones
of Eq. (\ref{A}) obtained by chosing the first qubit to play a special role.
The pair of four-vectors $({\xi}^{(T)}, {\eta}^{(T)})$ and $({\xi}^{(U)},{\eta}^{(U)})$ are defined accordingly. Moreover, using the dictionary Eq.(\ref{corres}), we can express the following results in terms of the dyonic charges.
Let us now define the following set of three  complex four vectors
\beq
\label{nvec}
n^{(TU)}_I=\begin{pmatrix}1\\T\\U\\TU\end{pmatrix},\quad
n^{(US)}_I=\begin{pmatrix}1\\S\\U\\US\end{pmatrix},\quad
n^{(ST)}_I=\begin{pmatrix}1\\S\\T\\ST\end{pmatrix}.
\eeq
\noindent
Notice that these are null with respect to our metric Eq.(\ref{expl}), i.e. $n\cdot n=0$ due to their tensor product structure (e.g. $n^{(TU)}=(1,U)^t\otimes (1,T)^t$).
With this notation the BPS mass can be written as
\beq
\label{BPStwist}
M^2_{BPS}=\frac{1}{8S_2T_2U_2}{\vert(S{\xi}^{(S)}-{\eta}^{(S)})\cdot n^{(TU)}\vert}^2,
\eeq
\noindent
where due to triality one can permute the labels $STU$ cyclically. 
Extremization with respect to $S$, $T$ and $U$ and their complex conjugates yields the equations

\beq
\label{extrequ1}
(\overline{S}{\xi}^{(S)}-{\eta}^{(S)})\cdot n^{(TU)}=0,
\eeq
\noindent
\beq
\label{extrequ2}
(\overline{T}{\xi}^{(T)}-{\eta}^{(T)})\cdot n^{(US)}=0,
\eeq
\noindent
\beq
\label{extrequ3}
(\overline{U}{\xi}^{(U)}-{\eta}^{(U)})\cdot n^{(ST)}=0
\eeq
\noindent
and their complex conjugates.

Let us use in the forthcoming manipulations the simplified notation ${\xi}_I\equiv{\xi}_I^{(S)}$ and ${\eta}_I\equiv{\eta}_I^{(S)}$  $I=1,2,3,4$. This means that in the following we look at the system of equations above from the viewpoint of the first qubit.
Substracting the conjugate of Eq.(\ref{extrequ3}) from Eq.(\ref{extrequ1})
we get
\beq
\label{useit}
(T-\overline{T})(U\overline{S}{\xi}_1-U{\eta}_1-\overline{S}{\xi}_3+{\eta}_3)=0
\eeq
\noindent
yielding for nonzero $T_2$ the equations
\beq
\label{US}
U=\frac{\overline{S}{\xi}_3-{\eta}_3}{\overline{S}{\xi}_1-{\eta}_1},\quad
\overline{S}=\frac{U{\eta}_1-{\eta}_3}{U{\xi}_1-{\xi}_3}.
\eeq
\noindent
Adding the conjugate 
of Eq.(\ref{extrequ3}) to Eq.(\ref{extrequ1})
and using Eq.(\ref{useit}) in the result we get
\beq
\label{useit2}
U\overline{S}{\xi}_2-U{\eta}_2-\overline{S}{\xi}_4+{\eta}_4=0
\eeq
\noindent
which implies
\beq
\label{US2}
U=\frac{\overline{S}{\xi}_4-{\eta}_4}{\overline{S}{\xi}_2-{\eta}_2},\quad
\overline{S}=\frac{U{\eta}_2-{\eta}_4}{U{\xi}_2-{\xi}_4}.
\eeq
\noindent
From Eqs.(\ref{US}) and (\ref{US2}) we see that
\beq
\label{nulldir1}
(\overline{S}{\xi}^{(S)}-{\eta}^{(S)})^2=0, \quad
(\overline{U}{\xi}^{(U)}-{\eta}^{(U)})^2=0.
\eeq
\noindent
Doing similar manipulations with the remaining equations (or which is the same permuting differently the labels $S$, $T$ and $U$) we obtain the constraint
\beq
\label{nulldir2}
(\overline{T}{\xi}^{(T)}-{\eta}^{(T)})^2=0.
\eeq
\noindent
The last two set of equations and their conjugates show that the six {\it complex} four-vectors appearing in Eqs.(\ref{extrequ1})-(\ref{extrequ3}) are {\it null}.
In the formalism of Section II. they define the principal null directions
for the planes in ${\bf C}^4$ spanned by the three pairs of four-vectors 
$({\xi}^{(S)},{\eta}^{(S)})$,
$({\xi}^{(T)},{\eta}^{(T)})$, and
$({\xi}^{(U)},{\eta}^{(U)})$.
Solving the quadratic equations we can write these principal null directions
in the form of Eq.(\ref{elso}) with appropriate labels $S$, $T$ or $U$ to be attached. Notice that in Eq.(\ref{elso}) $D$ is complex. Here the components of $\xi$ and $\eta$ are real and due to consistency we have to require that the quantity under the square root in Eq.(\ref{elso}) must be real and positive.
This ensures that the moduli are complex hence the K\"ahler potential is well defined.

Using Eq. (8), and the fact that the K\"ahler potential $e^{-K}=-8S_2T_2U_2$ should be positive\cite{Shmakova}, from the viewpoint of the first qubit the frozen values of the moduli are 
\beq
\label{frozenS}
S=\frac{(\xi\cdot\eta)-i\sqrt{-D}}{(\xi\cdot\xi)},\quad T=\frac{\overline{S}{\xi}_2-{\eta}_2}{\overline{S}{\xi}_1-{\eta}_1}, 
\eeq
\noindent
with the value of $U$ expressed in terms of $\overline{S}$ given by the first formula of Eq.(\ref{US}) or alternatively the first of Eq.(\ref{US2}).
These formulae imply that
\beq
\label{UT}
UT=\frac{\overline{S}{\xi}_4-{\eta}_4}{\overline{S}{\xi}_1-{\eta}_1}
\eeq
\noindent
providing the useful formula for $n^{(TU)}$  
\beq
\label{nTU}
n_I^{(TU)}=\frac{1}{\overline{S}{\xi}_1-{\eta}_1}(\overline{S}{\xi}_I-{\eta}_I).
\eeq
\noindent
Let us write this relation in the form
\beq
\label{alternate}
\frac{1}{\overline{S}{\xi}_1-{\eta}_1}\begin{pmatrix}\overline{S}{\xi}_1-{\eta}_1&\overline{S}{\xi}_2-{\eta}_1\\
\overline{S}{\xi}_3-{\eta}_3&\overline{S}{\xi}_4-{\eta}_4\end{pmatrix}=\begin{pmatrix}1\\U\end{pmatrix}\begin{pmatrix}1&T\end{pmatrix}.
\eeq
\noindent
Using this and its conjugate in Eqs.(\ref{matr1}-\ref{matr2}) we obtain
the transformed state Eq.(\ref{newstate}) as
\beq
\label{newstate2}
\vert{\psi}^{\prime}\rangle=\sqrt{\frac{2T_2U_2}{S_2}}\left[({\eta}_1-\overline{S}{\xi}_1)\vert 000\rangle -({\eta}_1-S{\xi}_1)\vert 111\rangle\right]
\eeq
\noindent
which is of the generalized GHZ form.
Hence we obtained the nice result: finding the frozen values of the moduli
for STU black holes is equivalent to finding an optimal distillation protocol for a GHZ state starting from the one defined by the charges as in Eq. (\ref{corres}).

In fact we can simplify further our expression for the transformed state as follows. First notice that the BPS mass is not sensitive to multiplication to an overall phase factor appearing in $\vert{\psi}^{\prime}\rangle$.
Moreover, a straightforward calculation shows that
\beq
\label{abs}
{\vert \overline{S}{\xi}_1-{\eta}_1\vert}^2=-\frac{1}{2}\frac{
({\xi}^{(T)}\cdot{\xi}^{(T)})
({\xi}^{(U)}\cdot{\xi}^{(U)})}
{({\xi}^{(S)}\cdot{\xi}^{(S)})}.
\eeq
\noindent
Recalling Eq. (\ref{frozenS}) we see that $S_2=-\sqrt{-D}/({\xi}^{(S)}\cdot{\xi}^{(S)})$ and similar expressions for $T_2$ and $U_2$ hold.
Collecting everything we get
\beq
\label{verynew}
\vert{\psi}^{\prime}\rangle =\vert D\vert^{1/4}(\vert 000\rangle +e^{i\delta}\vert 111\rangle),
\eeq
\noindent
where
\beq
\label{fazis}
\delta=\pi+2\arctan\frac{S_2{\xi}_1}{S_1{\xi}_1-{\eta}_1}.
\eeq
\noindent
Notice that this state is of the
Eq. (\ref{slocckan}) form, verifying the claim that the process of finding the frozen values for the moduli for BPS STU black holes is equivalent to finding the canonical form of the corresponding three qubit state using complex SLOCC transformations.

Having the exact values of ${\psi}_{000}^{\prime}$ and ${\psi}_{111}^{\prime}$ at our disposal we can put these into our formula Eq. (\ref{result}) yielding the extremal value for the BPS mass the expression $M^2_{BPS}{\vert}_{\rm extr}=\vert Z\vert^2_{\rm extr}=\vert Z\vert^2_{\rm hor}=\sqrt{-D}$.
Hence according to Eqs. (\ref{entropy}) and (\ref{alternative}) our final result for the macroscopic entropy of the extremal STU BPS black hole is
\beq
S_{BH}=\pi\sqrt{-D}=\frac{\pi}{2}\sqrt{{\tau}_{ABC}}
\eeq
\noindent
where using the correspondence between our three-qubit amplitudes and dyonic charges Eq.(\ref{corres}) we obtain
\begin{widetext}
\beq
-D=({\xi}\cdot{\xi})(\eta\cdot\eta)-(\xi\cdot\eta)^2=-(p\circ q)+4((p^1q_1)(p^2q_2)+(p^1q_1)(p^3q_3)+(p^2q_2)(p^3q_3))-4p^0q_1q_2q_3+4q_0p^1p^2p^3,
\eeq
\end{widetext}
\noindent
where
$p\circ q=p^0q_0+p^1q_1+p^2q_2+p^3q_3$.
This expression for the black hole entropy expressed in terms of the charges has been obtained in Ref. [18] by solving the stabilization equations Eq. (\ref{stabi}). Comparing our results Eqs.(\ref{US}) and (\ref{frozenS}) for the frozen values of the moduli with that paper (see the somewhat more complicated looking expressions as given by Eqs. (32) and (35) of Ref. [18] we find that they agree. (Recall, however our different sign convention, see the paragraph following Eq. (\ref{corres}.) 
The observation that the black hole entropy can be expressed in the nice form as the negative of
Cayley's hyperdeterminant is due to Duff\cite{Duff}.
Here we presented a complete rederivation of this result 
using the language of quantum information theory.
Our approach also provided some new insights into this process in the form of Eqs. (\ref{newstate}), (\ref{SLOCC}) and (\ref{verynew}).
These expressions show, that the process of finding the frozen values of the moduli is equivalent to the quantum information theoretic one of performing
an optimal set of SLOCC transformations on the initial three-qubit state with integer amplitudes to arrive at a state of GHZ form.
It is important to realize however, that though the transformed states
seem to be complex, they are really real states in disguised form.
This means that the invariant $\Delta$ of Eq. (\ref{Delta}) is vanishing for both of these states so according to the results of Section II. we can find a basis where they are real. 
In the next section we explore a little bit more the geometric meaning of these reality conditions and the embedding of the real entangled states
of the STU model in the most general complex ones of three-qubit entanglement.

\section{A geometric classification of STU Black Holes}

In the previous section we considered double extreme BPS STU black holes.
We concluded that these black holes are characterized by the constraints
${\tau}_{ABC}\neq 0$ and $\Delta =0$, meaning that in the SLOCC classification
these black holes are in the subclass of the GHZ class,  characterized by a special reality condition.
As we already know there are {\it two} different classes of real states in the GHZ class characterized by the conditions (\ref{cond1}) and (\ref{cond2}).
The second of these conditions means that the principal null directions as four vectors in ${\bf C}^4$ are complex conjugate of one another.
These conditions characterize the BPS STU black holes.
What about the other condition?

In the paper of Kallosh and Linde\cite{Linde} the authors using the SLOCC classification of three-qubit states presented a complete classification of STU black holes.
The black holes corresponding to the GHZ class are called "large" black holes.
This term means that these black holes have classically non vanishing event horizons. According to that authors there are two different classes of such black holes. One of them is the BPS black hole studied in the previous section. 
The other class corresponds to large non-BPS black holes.
It is easy to demonstrate that these black holes are characterized by the other set of reality conditions namely the one of Eq. (\ref{cond1}).
Indeed, according to this condition such states have two linearly independent {\it real} four vectors as their principal null directions.
For example the canonical GHZ state $\vert 000\rangle +\vert 111\rangle$ belongs to this class.
It should be also clear by now that the two different reality classes can also
be characterized by the {\it sign} of the Cayley hyperdeterminant.
Positive sign corresponds to non-BPS and negative to BPS black holes.
Our observations based on the reality conditions Eqs. (\ref{cond1}) and (\ref{cond2}) can be regarded as a refinement of the classification of Ref. [13], by also clarifying the embedding of these entangled states corresponding to "large" STU black hole solutions into the complex states of more general type used in quantum information theory.

Next the classification of Ref. [13] proceeds also to include
the so called "small" black holes. These are the ones with classically vanishing horizons corresponding to the vanishing of ${\tau}_{ABC}$.
These black holes are  represented by the separable classes, and the W-class (see Section II.)
In the following we show that using the language of twistor theory we can obtain a nice geometric characterization of this classification.

The basic objects of our geometric correspondence are pairs of complex four vectors. These are elements of the twistor space ${\bf C}^4$.
These pairs of complex four vectors span {\it planes} in ${\bf C}^4$.
Since our coordinates are defined merely projectively, it is convenient to switch to the projective picture and use the projective twistor space which is
${\bf CP}^3$. In this space our pairs of complex four-vectors define {\it complex lines}.
For example our four-vectors ${\xi}^{(S)}$ and ${\eta}^{(S)}$ with integer components  used in the previous section for an arbitrary complex number $S$ define the line $S{\xi}^{(S)}-{\eta}^{(S)}$ in ${\bf CP}^3$. Alternatively for complex $T$ and $U$ we can also define the lines $T{\xi}^{(T)}-{\eta}^{(T)}$ and
$U{\xi}^{(U)}-{\eta}^{(U)}$.
Explicitly we have
\beq
\label{SS}
{\xi}^{(S)}_I=\begin{pmatrix}p^0\\p^2\\p^3\\q_1\end{pmatrix},\quad
{\eta}^{(S)}_I=\begin{pmatrix}p^1\\q_3\\q_2\\-q_0\end{pmatrix},
\eeq
\beq                                                                            {\xi}^{(T)}_I=\begin{pmatrix}p^0\\p^1\\p^3\\q_2\end{pmatrix},\quad              {\eta}^{(T)}_I=\begin{pmatrix}p^2\\q_3\\q_1\\-q_0\end{pmatrix},                 \eeq
\beq                                                                            {\xi}^{(U)}_I=\begin{pmatrix}p^0\\p^1\\p^2\\q_3\end{pmatrix},\quad              {\eta}^{(S)}_I=\begin{pmatrix}p^3\\q_2\\q_1\\-q_0\end{pmatrix}.                 \eeq
\noindent
Of course these lines are very special compared to the ones of most general complex type.

Let us describe three-qubit entanglement of the most general type from the viewpoint of one of the parties e.g. Alice. The eight complex amplitudes characterizing this type of entanglement are then characterized by the vectors ${\xi}^{(A)}$and ${\eta}^{(A)}$ of Eq. (\ref{A}).  
The important special case related to black holes is obtained by restricting these complex amplitudes to ${\xi}^{(S)}$ and ${\eta}^{(S)}$ i.e. to the ones of Eq. (\ref{SS}). In the following we drop the superscript $(A)$ or $(S)$ again to reduce clutter in notation.
Let us now define a nondegenerate quadratic form $Q: {\bf C}^4\times {\bf C}^4\to{\bf C}$ as follows. For $\xi,\eta\in {\bf C}^4$ define
\beq
\label{quadric}
Q(\xi,\eta)\equiv\xi\cdot\eta = {\xi}_1{\eta}_4-{\xi}_2{\eta}_3-{\xi}_3{\eta}_2+{\xi}_4{\eta}_1.
\eeq
\noindent
Then the vectors $\zeta\in {\bf C}^4$ satisfying $Q(\zeta,\zeta)=0$ define a quadric surface ${\cal Q}$ in ${\bf CP}^3$.
We regard the twistor space with this quadric ${\cal Q}$ as fundamental.

Let us now consider an arbitrary complex line corresponding to a three-qubit state in ${\bf CP}^3$ of the form
$w{\xi}-{\eta}$ where $w$ is a nonzero complex number and ${\xi}$ and $\eta$ are non null i.e. they are not lying on the quadric ${\cal Q}$.
In the following we shall examine the intersection properties of a complex line
of the above form with the fixed quadric ${\cal Q}$.
When the equation $Q(w\xi-\eta,w\xi-\eta)=0$ has two solutions for $w$ (corresponding to the two principal null directions)
the line intersects ${\cal Q}$ in two different points.
The sufficient and necessary condition for this to happen is just $D\neq 0$
i.e. ${\tau}_{ABC}\neq 0$.
Hence states belong to the GHZ class iff the representative lines intersect ${\cal Q}$ in two points.
Large black holes within this class are represented by the {\it real} lines
described by the vectors of Eq. (\ref{SS}) with integer components.
They are either BPS ($D<0$) or non BPS ($D>0$).
In the first case the principal null directions defined by the frozen value of the moduli $S$ on the horizon $u^{\pm}$ are complex conjugate of one another, in the other they are real (see Fig. 1.).

\begin{figure}
\centerline{\resizebox{5.0cm}{!}{\includegraphics{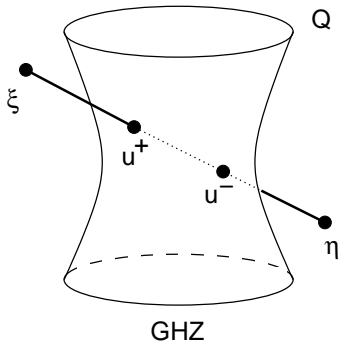}}}
\caption{\label{fig1}
Geometric representation of large black holes corresponding to real states in   the GHZ class. The line is defined by the vectors $\xi$ and $\eta$ of Eq. (86) defined by the dyonic charges. The points of intersection of the line with the quadric ${\cal Q}$ are the principal null directions $u^{\pm}$ defined by the frozen value of the moduli $S$ Eq. (76) on the horizon.} 
 \end{figure}
 
If the equation $Q(w\xi-\eta,w\xi-\eta)=0$ has merely one solution (the case of one principal null direction) the line is tangent to the quadric ${\cal Q}$
at this particular point. This can happen iff $D=0$ i.e. ${\tau}_{ABC}=0$.
Then states belong to the W-class iff the corresponding lines are tangent to the quadric.
After specializing again to real states now representing the small black holes
we obtain the geometric situation depicted by Fig. 2.

\begin{figure}
\centerline{\resizebox{5.0cm}{!}{\includegraphics{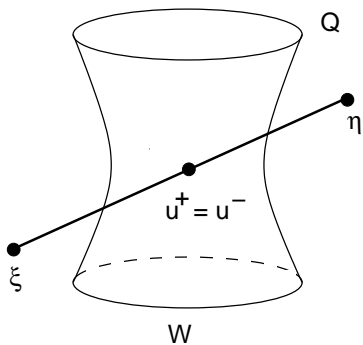}}}
\caption{\label{fig2}
Geometric representation of small black holes corresponding to real states in   the
 W class. The line is defined by the vectors $\xi$ and $\eta$ defined by the
dyonic charges. The point of intersection of the line tangent to the quadric    ${\cal Q
}$ corresponds to the two coincident principal null directions $u^{\pm}$.} 
 \end{figure}
Note, however that in these two cases of genuine three-qubit entanglement the points through which the lines were defined are themselves {\it not} lying on ${\cal Q}$.

The next special case is the one of $A(BC)$ separable states.
In this case ${\tau}_{A(BC)}=0$ hence according to Eq. (\ref{conc2})
the vectors $\xi$ and $\eta$ are proportional, hence our line degenerates to a point not lying on the quadric ${\cal Q}$. Including also the degenerate case
when one of the vectors e.g. $\eta$ is vanishing, we can represent the
corresponding situation of small black holes by drawing a point off the quadric
represented by the vector $\xi$ now with integer components (see Fig. 3).
                                                                                \begin{figure}                                                                  \centerline{\resizebox{5.0cm}{!}{\includegraphics{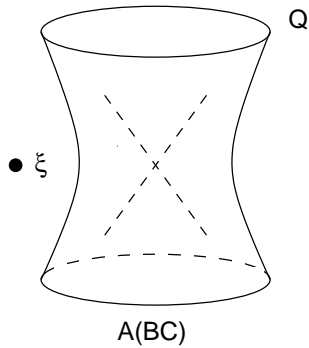}}}                 \caption{\label{figr3}                                                           Geometric representation of small black holes corresponding to real states in the A(BC) class. The point off the quadric ${\cal Q}$ is defined by the vector $\xi$ of dyonic charges. The other vector $\eta$ is either projectively equivalent to $\xi$ or vanishing.
The dashed lines intersecting at a point refer to the existence of two families of lines on ${\cal Q}$ ruling it.} \end{figure}

Let us now turn to the cases when the lines themselves are lying inside the quadric ${\cal Q}$. Such lines are called {\it isotropic}\cite{Hughston} with respect to ${\cal Q}$. It is well-known that there are exactly two families of lines
 on a nondegenerate quadric ${\cal Q}$ in ${\bf CP}^3$.  In other words our quadric ${\cal Q}$ is {\it ruled} by two families of lines. They are conventionally called $\alpha$-lines and ${\beta}$-lines\cite{Hughston}. 
Two of such representative lines are depicted in Fig. 3.
Two lines belonging to the same family do not intersect; whereas, two lines
belonging to the opposite families intersect at a single point (see Fig. 3.) on ${\cal Q}$.
Hence any nondegenerate quadric in ${\bf CP}^3$ is isomorphic to ${\bf CP}^1\times{\bf CP}^1$.
Using the results of our previous paper\cite{Levay1} one can show that these isotropic lines correspond precisely to $B(AC)$ and $C(AB)$ separable states.
In order to see this recall that\cite{Levay1} by defining ${\tau}_+={\tau}_{C(AB)}$ and ${\tau}_-={\tau}_{B(AC)}$ we have
\beq
\label{jaj}
{\tau}_{\pm}=\vert\xi\cdot\xi\vert^2+2\vert\xi\cdot\eta\vert^2+\vert\eta\cdot\eta\vert^2+(P^{IJ}\mp\ast P^{IJ})\overline{P}_{IJ},
\eeq
\noindent
where
\beq
\ast P_{IJ}\equiv \frac{1}{2}{\varepsilon}_{IJKL}P^{KL},
\eeq
\noindent
and see Eq. (\ref{Plucker}) for the definition of the Pl\"ucker coordinates.
Isotropic lines satisfy the relations $\xi\cdot\xi=\eta\cdot\eta=\xi\cdot\eta=0$, moreover such lines are necessarily self-dual or anti-self-dual\cite{Hughston}. Hence for isotropic lines we have either ${\tau}_+=0$ or ${\tau}_-=0$.
Conversely, using the positivity\cite{Levay1} of the terms in Eq. (\ref{jaj}) the vanishing of ${\tau}_{\pm}$ implies that the corresponding lines are isotropic.
Since the states are C(AB) or B(AC) separable if and only if ${\tau}_+$ or ${\tau}_-$ vanishes, isotropic lines on $ {\cal Q}$ represent precisely two of our biseparable classes.
Specializing again to real states of $C(AB)$ or $B(AC)$ separable form
representing small STU black holes we have the geometrical situation of Figs. 4 and 5. 

\begin{figure}
\centerline{\resizebox{5.0cm}{!}{\includegraphics{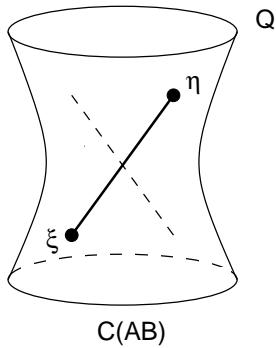}}}
\caption{\label{figr4}                                                           Geometric representation of small black holes corresponding to real states in
the $C(AB)$ class.
The line through the points $\xi$ and $\eta$ lying now on ${\cal Q}$ is an      isotropic
line, i.e. it lies entirely inside the quadric and coincides with one from the family of special lines of ${\cal Q}$. These lines are related to the self-duality of the Pl\"ucker matrix and are called $\alpha$-lines.
} \end{figure}

\begin{figure}
\centerline{\resizebox{5.0cm}{!}{\includegraphics{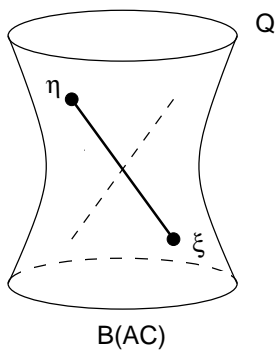}}}
\caption{\label{figr5}
Geometric representation of small black holes corresponding to real states in
the $B(AC)$ class.
The line through the points $\xi$ and $\eta$ lying now on ${\cal Q}$ is an 
isotropic line, coinciding with one from the other
family of special lines of ${\cal Q}$. These lines are related to the anti-self-duality
of the Pl\"ucker matrix and are called $\beta$-lines.
} \end{figure}

Finally we are left with the geometrical representation of the small black holes corresponding to the totally separable class, i.e. the states of the form   
$(A)(B)(C)$.
Such states are represented by points since they are $A(BC)$ separable, moreover they have to lie on the quadric since due to $C(AB)$ and $B(AC)$ separability they are parts of isotropic lines. The only possible way of representing them is
by a point on the quadric which is of course located at the intersection of
an $\alpha$ and a $\beta$-plane (see Fig. 6).

\begin{figure}
\centerline{\resizebox{5.0cm}{!}{\includegraphics{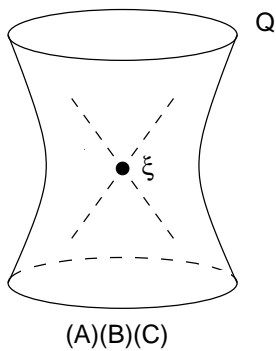}}}
\caption{\label{figr6}
Geometric representation of small black holes corresponding to real states in
the totally separable $(A)(B)(C)$ class.
Such holes are represented by a single point lying at the intersection of an $\alpha$ and a $\beta$-line. Here this point is represented by $\xi$. The other point ${\eta}$ is either projectively equivalent to ${\xi}$ or can be taken to be zero.  
} \end{figure}

Note that this geometrical representation is from the viewpoint of system $A$
or which is the same the $S$-part of the STU model. 
The fixed quadric ${\cal Q}$ is defined by using the $SL(2,{\bf R})\otimes SL(2, {\bf R})$ invariant metric Eq. (\ref{expl}).
Since the symmetry of the STU model is $[SL(2,{\bf R})/SO(2)]^{\otimes 3}$
(the moduli are coordinates of this manifold)
this choice of ${\cal Q}$ is dictated by the basic structure of the STU model.
Physically however, all parties are equivalent hence the geometric picture as given by Figs. 1-6. is independent from the choice of parties.
We can give however, to the subsystem $A$ a {\it physically} different role 
by allowing transformations on the combined system $BC$ (i.e. $TU$) of more general type. For example instead of applying the real version of the SLOCC group $SL(2, {\bf R})^{\otimes 3}$ we can have the larger one $SL(2, {\bf R})\otimes SO(2,2)$. This means that $B$ and $C$ are sharing among each other local resources of
a more general kind than $A$.
This enlargement of the SLOCC group in the entanglement picture amounts to
using a dual description of black holes where the moduli $S$ is singled out
and whose imaginary part plays the role of the string coupling constant\cite{Rahmfeld, Shmakova, Kallosh}.
The manifold for the moduli in this picture is $\frac{SL(2, {\bf R})}{SO(2)}\times\frac{SO(2,2)}{SO(2)\times SO(2)}$.
The different roles the parties $A$ and $BC$ play in the local protocols  
performed by them corresponds to the different characters $S$-duality and
$T$ and $U$ dualities have in string theory.
Indeed $S$-duality ( associated with the $SL(2, {\bf Z})$ subgroup of $SL(2, {\bf R})$) in this picture is of nonperturbative whereas $T$ and $U$ dualities based on $SO(2,2)$ symmetry are of perturbative character (i.e. they are not mixing electric and magnetic charges).
This point has been emphasized by Kallosh and Linde\cite{Linde}.

In our geometric representation this dual picture means that now Figs 1-6.
repesent {\it the} physical situation of $S(TU)$ black holes.
Althoug the nondegenerate quadric ${\cal Q}$ is now represented differently (i.e. in the $SO(2,2)$ form) the intersection properties are invariant.
The choice of base describing the situation is the one obtained after applying
an $Sp(8, {\bf R})$ transformation to the charges\cite{Shmakova} which is in our labelling of the three-qubit system is equivalent to the $O(4, {\bf R})$ transformation
\beq
\label{dualp}
\hat{p}^I=\begin{pmatrix}\hat{p}^1\\\hat{p}^2\\\hat{p}^3\\\hat{p}^4\end{pmatrix}
=\frac{1}{\sqrt{2}}\begin{pmatrix}{\xi}_1-{\xi}_4\\{\xi}_2+{\xi}_3\\-{\xi}_1-{\xi}_4\\-{\xi}_2+{\xi}_3\end{pmatrix}
\eeq

\beq
\label{dualq}
\hat{q}^I=\begin{pmatrix}\hat{q}^1\\\hat{q}^2\\\hat{q}^3\\\hat{q}^4\end{pmatrix}
=\frac{1}{\sqrt{2}}\begin{pmatrix}{\eta}_1-{\eta}_4\\{\eta}_2+{\eta}_3\\-{\eta}_1-{\eta
}_4\\-{\eta}_2+{\eta}_3\end{pmatrix},
\eeq
\noindent
where now indices are lowered with the $SO(2,2)$ invariant metric $h_{IJ}={\rm diag}(-1,-1,+1,+1)$. 
Clearly $\hat{p}^I\hat{q}_{I}=h_{IJ}\hat{p}^I\hat q{J}=\xi\cdot\eta$, see
Eq. (\ref{quadric}).
In closing this section we note that the choice of base Eqs. (\ref{dualp}) and (\ref{dualq}) corresponds to the real version of the so called magic base of Hill and Wootters \cite{Wootters}, which is related to the usual conversion of four-vector indices to spinorial ones of twistor theory\cite{Penrose}.

\section{Conclusions}

In this paper we have studied intresting similarities between two different fields of theoretical physics, quantum information theory and the physics of stringy black holes.
Though they are seemingly unrelated, one can realize that the unifying themes in both of these fields such as
information, entropy, and entanglement are the same.
Since the near horizon geometry of black holes is $AdS^2\times S^2$ using the idea of Ads/CFT holography one might certainly expect connections between entanglement entropy and black hole entropy. 
Though there are some interesting recent developments\cite{Brustein} in relating entanglement entropy and black hole entropy, the correspondence between these notions is not well-understood.
In order to get some further insight into the nature of such important problems it is sometimes useful to look for the clues
coming from different strains of knowledge.
Hence, following the initiative of Duff\cite{Duff} and Kallosh and Linde \cite{Linde}
in the present paper we have established new relations between extremal black holes in the $STU$-model of string theory and qubit systems in quantum information theory.

In particular we have shown that the well-known process of finding the frozen values for the moduli on the horizon in the theory of STU black holes corresponds to the problem of finding a canonical form for the three qubit state
defined by the dyonic charges using ${\it complex}$ SLOCC transformations in quantum information theory.
Alternatively, this process equivalent to solving the stabilization (attractor)
equations in one picture corresponds to obtaining the optimal distillation protocol for a  GHZ-state in the other.
The geometric representation for this process was found. It is equivalent to finding the principal null directions of a complex plane in ${\bf C}^4$. 
We have managed to characterize geometrically the real states describing STU black holes by embedding them inside the more general complex ones 
used in quantum information theory.
Using the language of twistors based on the intersection properties of complex lines with a fixed quadric ${\cal Q}$ in ${\bf CP}^3$
an instructive geometric classification for STU black holes
was given.

Let us now add some important observations to these results.
Let us first consider the transformed state of Eq. (\ref{newstate}).
As we have shown using the frozen values for the  moduli $S$, $T$ and $U$
results in the state of the $GHZ$ form Eq. (\ref{verynew}). 
Since the amplitudes of this state besides  ${\psi}^{\prime}_{000}$ and ${\psi}^{\prime}_{111}$ are zero
the projection onto these components in Eq. (\ref{result}) is not needed.
Hence $M^2_{BPS}\vert_{\rm extr}=\frac{1}{2}{\vert\vert{\psi}^{\prime}\vert\vert^2}_{\rm hor}=\frac{1}{2}\langle\psi\vert C_U^{\dagger}C_U\otimes B_T^{\dagger}B_T\otimes A_S^{\dagger}A_S\vert\psi\rangle_{\rm hor}$. 
Then we get for the black hole entropy
\beq
\label{norma}
S_{BH}=\frac{\pi}{2}{\vert\vert{\psi}^{\prime}\vert\vert^2}_{\rm hor}=\frac{\pi}{2}\sqrt{{\tau}_{ABC}}.
\eeq
\noindent
This interesting formula relates the black hole entropy to the value of the norm of the transformed state at the horizon.
Now in papers\cite{Acin3,Verstr} the optimal local distillation protocol for the {\it canonical} GHZ state $\frac{1}{\sqrt{2}}(\vert 000\rangle +\vert 111\rangle)$ was found.
In particular it was proved\cite{Verstr} that the total probability for obtaining the canonical GHZ state is bounded from above by $\sqrt{{\tau}_{ABC}}/{\lambda}_{max}(C^{\dagger} C\otimes B^{\dagger}B\otimes A^{\dagger} A)$. 
Here ${\lambda}_{max}(X)$ denotes the largest eigenvalue of the operator $X$ and the parameter dependent operators $A, B, C$ are the generalizations of our $A_S, B_T, C_U$ of Eq. (\ref{SLOCC})  for the complex case.
Hence an upper bound is achieved by minimizing this largest eigenvalue with respect to the parameters.
These observations show that in the case of BPS STU black holes the minimum area principle of the supersymmetric attractor mechanism is somehow related to the maximization of the probability of success
for converting a particular state to the canonical GHZ state .
It would be interesting to use the insight and formalism provided by stringy black holes for obtaining an alternative description  
of this optimization process. 

As our second observation let us consider the {\it real} state
\beq
\label{realtrafo}
\vert\hat{\psi}(S,T,U)\rangle\equiv {\cal C}_U\otimes {\cal B}_T\otimes {\cal A}_S\vert \psi\rangle,
\eeq
\noindent
known from Eq. (\ref{elsotrukk}).
Then it is easy to show that
\beq
\label{formation1}
\langle\hat{\psi}\vert{\sigma}_2\otimes{\sigma}_2\otimes I\vert\hat{\psi}\rangle={\rm Tr}(\hat{\varrho}_{BC}{\sigma}_2\otimes{\sigma}_2),
\eeq
\noindent
where  now $\hat{\varrho}_{BC}(S,T,U)\equiv\hat{\psi}_{lk0}\hat{\psi}_{l^{\prime}k^{\prime}0}+\hat{\psi}_{lk1}\hat{\psi}_{l^{\prime}k^{\prime}1}=\hat{\xi}_I\hat{\xi}_J+\hat{\eta}_I\hat{\eta}_J$ (compare with Eq. (\ref{BCsur})).
Using similar manipulations for the expectation values of the operators
${\sigma}_2\otimes I\otimes {\sigma}_2$ and $I\otimes{\sigma}_2\otimes{\sigma}_2$ we obtain for the BPS mass squared Eq. (\ref{elsotrukk}) the formula
\begin{widetext}
\beq
\label{formation2}
M_{BPS}^2=\frac{1}{8}(\vert\vert\hat{\psi}\vert\vert^2+{\rm Tr}(\hat{\varrho}_{BC}{\sigma}_2\otimes{\sigma}_2)+
{\rm Tr}(\hat{\varrho}_{
AC}{\sigma}_2\otimes{\sigma}_2)+
{\rm Tr}(\hat{\varrho}_{
AB}{\sigma}_2\otimes{\sigma}_2)).
\eeq
\end{widetext}
\noindent
Now in Ref. 17. it was shown that the magnitudes 
${\cal C}_{BC}\equiv\vert{\rm Tr}(\hat{\varrho}_{BC}{\sigma}_2\otimes{\sigma}_2)\vert$ etc. define the {\it concurrences} for the real qubits i.e. rebits. Moreover, this quantity defines the important 
quantity, {\it the entanglement of formation} for rebits via the formula
\beq
E(\hat{\varrho}_{BC})=H\left(\frac{1+\sqrt{1-{\cal C}^2( \hat {\varrho}_{BC} )  }}{2} \right)  
\eeq
\noindent
where
$H(x)=-x\log_2x-(1-x)\log_2(1-x)$ is the binary Shannon entropy.
Since $\vert\hat{\psi}\rangle$ and $\vert{\psi}^{\prime}\rangle$ are unitarily related (see Eq. (\ref{diag})) we have 
$\vert\vert\hat{\psi}\vert\vert^2=\vert\vert{\psi}^{\prime}\vert\vert^2$
hence the extremal BPS mass squared can also be written in the form
${M^2_{BPS}}\vert_{\rm extr}=\frac{1}{2}\overline{\cal C}_{\rm hor}$ where
$\overline{\cal C}=\frac{1}{3}({\cal C}_{AB}+{\cal C}_{BC}+{\cal C}_{AC})$
is the average {\it real} concurrence.
Hence the entropy for the large BPS STU black hole can be written in the alternative forms
\beq
\label{altentropy}
S_{BH}=\frac{\pi}{2}\sqrt{ \hat{\tau}_{ABC} }=\frac{\pi}{2}\overline{\cal C}_{\rm hor}=\frac{\pi}{2}\vert\vert\hat{\psi}\vert\vert^2_{\rm hor}.      
\eeq
\noindent
Notice that in these expressions all quantities are expressed in terms of the real moduli dependent three-qubit state $\vert\hat{\psi}\rangle$ Eq. (95) calculated with  the frozen values for them at the horizon. 
Of course due to the $SL(2,{\bf R})$ invariance of the three-tangle we have
$\hat{\tau}_{ABC}={\tau}_{ABC}$ so it has the same value, no matter we use the state $\vert\psi\rangle$ with integer or the one $\vert\hat{\psi}\rangle$ with moduli dependent real amplitudes. However, the norm and the average real concurrence depends on the values of the moduli in a nontrivial way. Indeed, according to
Eq. (\ref{formation2}) the combination of these quantities gives $M_{BPS}^2$ to be extremized.
However, quite remarkably all three quantities are frozen to the same value
at the horizon.

The occurrence of the real concurrence (of which the real entanglement of formation is a monotonically increasing and convex function) in the STU black-hole scenario suggests a possibility for an alternative physical interpretation of the macroscopic black hole entropy.
As it is well-known (see Ref. 34 for a nice review) the entanglement of formation of a two-qubit {\it mixed state} $\varrho$ is related to the {\it minimum} number of EPR pairs required to create that state $\varrho$.
More precisely we have the following definition. Let us consider {\it all}
pure state decompositions of the mixed state $\varrho$ of two qubits say $A$ and $B$ in the form
$\varrho=\sum_{k=1}^Mp_k\vert{\Phi}_k\rangle\langle{\Phi}_k\rangle$.
Let us moreover introduce the quantity $E(\Phi)=S({\rm Tr}_B\vert\Phi\rangle\langle\Phi\vert)=S({\rm Tr}_A\vert\Phi\rangle\langle\Phi\vert)$ with $S$ denoting the von Neumann entropy. Then the definition of the entanglement of formation is\cite{Wootters,Wootters2} 
\beq
\label{entanglement}
E(\varrho)={\rm inf}\sum_kp_kE({\Phi}_k),
\eeq
\noindent
where the infimum is taken over all pure state decompositions of $\varrho$.
These definitions and Eq. (\ref{altentropy}) clearly shows the possibility of relating the BPS STU black hole entropy to the {\it minimization} of the number of EPR pairs needed
to create a state characterized by the density matrices $\hat{\varrho}_{AB}$,
$\hat{\varrho}_{BC}$ and $\hat{\varrho}_{AC}$ as a function of the moduli fields.
This number according to the very definition of the entanglement of formation Eq. (\ref{entanglement}) is also related to the minimization of the convex hull of the von-Neumann entropies with respect to all possible pure state decompositions of the state ${\varrho}$. This idea relating the average entanglement of formation to the black hole entropy might turn out to be relevant in identifying the black hole entropy with the entanglement entropy within the framework of AdS/CFT correspondence. 

It is important to interpret the message of these sentences correctly.
The entanglement present in the physics of STU black holes is of unusual type.
Here the entanglement is {\it not} carried by distinguishable particles as in quantum information theory, but rather by special nonlocal objects that are composites of quantized charges and the moduli (see Eq. (95)).
Indeed the real entangled state of Eq. (95) is represented by an {\it entire line} in our geometric representation. Then when we are talking about entanglement of formation using EPR pairs etc. one has to have in mind this strange kind of entanglement. Of course according to the microscopic interpretation of black hole entropy since the quantized charges are related\cite{Linde} to the numbers of $D0$, $D2$, $D4$, and $D6$ branes  this kind of entanglement somehow should boil down to the
usual one of string theory states.

Returning back to the real concurrence, we stress that its square is {\it not}
the same as the restrictions of the squares of the complex concurrences , i.e. the quantities ${\tau}_{AB}$, ${\tau}_{BC}$ and ${\tau}_{AC}$ to the real domain.
In fact it is easy to show that for BPS STU black holes i.e. $D<0$ we have
\beq
\label{BPShole}
{\cal C}^2_{AB}={\tau}_{AB}+{\tau}_{ABC}.
\eeq
\noindent
However, for non-BPS STU black-holes i.e. $D>0$ the two concepts turn out to be identical, i.e. in this case we have for example ${\cal C}^2_{AB}={\tau}_{AB}$.
Looking back at the form of our reality conditions Eqs. (\ref{cond1}) and (\ref{cond2}) it is clear that using the notion of the real concurrence these expressions can be cast into a unified form. 
For example the reality condition for BPS STU black holes is ${\sigma}_{ABC}={\cal C}_{AB}{\cal C}_{BC}{\cal C}_{AC}$.

These considerations and the geometric representation of Section IV.
shows that the three-qubit states relevant to STU black holes are described by real {\it lines} in ${CP}^3$. These lines are lying on $SU(2)^{\otimes 3}$ orbits of the ones determined by the vectors ${\xi}$ and ${\eta}$ defined by the dyonic charges or of the more general ones determined by the ones $\hat{\xi}$ and ${\hat{\eta}}$ corresponding to the moduli dependent real states $\vert\hat{\psi}\rangle$.
In this paper we have used the complex geometry of three-qubit states of the most general type. However, the three-qubit states having some relevance for stringy black holes are at most real. Though we have clarified how these states are embedded in the space of the most general three-qubit ones 
the question arises, is there any relevance to string theory the existence of complex three-qubit states of the most general kind?
Though we do not know the answer, we note that the situation is somewhat similar to the one in twistor theory.
In twistor theory\cite{Penrose,Ward,Hughston} real lines (defined differently than here) in ${\bf CP}^3$ correspond to points of conformally compactified Minkowski space time, however to see the full power of  twistor geometry one is forced to include also {\it complex} lines corresponding to points of {\it complexified} and compactified Minkowski space time.
This Klein correspondence where instead of lines in ${\bf CP}^3$ points in a space (the complex Grassmannian Gr(2,4)) isomorphic to compactified and            complexified Minkowski space-time can be used to obtain a geometrical           representation for three-qubit states                                           similar to the one presented here\cite{Levay1}.
Notice that this correspondence between lines and space-time points is a non local one, which according to the original motivation of twistor theory is expected to play an important role in describing the nonlocality of quantum entanglement.

Though the similarity between the real lines found here and the ones of twistor theory is obvious it is not at all clear how can we relate these geometric considerations to the underlying special geometry of $N=2$, $D=4$
supergravity or to string theory states with some number of $D$-branes. Note, however that the special role of real coordinates (the analogue of our real lines found here) 
in supergravity theories is currently under investigation\cite{Mohaupt,Macia}.
For the moment the status of the new relations found in this paper is just like the ones of Refs. [10,13] that they are merely mathematical coincidences.
Though we are aware that the appearance of a mathematical structure in two disparate subjects does not necessarily imply a deeper unity, the realization that these relations do exist might turn out to be important for obtaining further insights for both string theorists and researchers working in the field of quantum information.

                                                                                \section{Acknowledgements}                                                      Financial support from the Orsz\'agos Tudom\'anyos Kutat\'asi Alap (OTKA),      (grant numbers T032453 and T038191) is gratefully acknowledged.                 We also thank Dr. L\'aszl\'o Borda for his help with the figures.

\end{document}